\documentclass[%
 reprint,
superscriptaddress,
 amsmath,amssymb,
 aps,
pra,
floatfix,
]{revtex4-2}
\usepackage{graphicx}
\usepackage{dcolumn}
\usepackage{bm}
\usepackage{color}
\usepackage{physics}
\usepackage{comment}
\usepackage{color}

\begin{document}

\preprint{APS/123-QED}

\title{High-rate qutrit entanglement swapping with photon-number-basis}

\author{Kazufumi Tanji}
\email{kizehemu@keio.jp}
\affiliation{%
 Department of Electronics and Electrical Engineering, Faculty of Science and Engineering, Keio University, Yokohama, Kanagawa 223-8522, Japan
}%

\author{Hikaru Shimizu}
\affiliation{%
 Department of Electronics and Electrical Engineering, Faculty of Science and Engineering, Keio University, Yokohama, Kanagawa 223-8522, Japan
}%

\author{Masahiro Takeoka}
\affiliation{%
 Department of Electronics and Electrical Engineering, Faculty of Science and Engineering, Keio University, Yokohama, Kanagawa 223-8522, Japan
}%
\affiliation{Advanced ICT Research Institute, National Institute of Information and
Communications Technology (NICT), Koganei, Tokyo 184-8795, Japan}
\date{\today}

\begin{abstract}
Entanglement generation between distant nodes is a fundamental process in distributed quantum information processing. Qudits, high-dimensional quantum states, are promising candidates for enhancing entanglement distribution capabilities. However, the success probability of qudit entanglement distribution using Bell measurements is typically lower than that of conventional qubit-based protocols. In this paper, we propose a novel entanglement swapping protocol specifically designed for qutrits (three-dimensional quantum states). Our protocol employs photon-number encoding combined with an additional mode basis, such as polarization, effectively increasing the success probability. We show that the proposed qutrit protocol can achieve higher entanglement generation rates compared to conventional two-photon detection-based qubit protocols, especially under conditions of limited photon generation probability. Furthermore, we evaluate our protocol under realistic experimental imperfections, including photon loss and threshold detection, and show that it achieves high fidelity with probabilistic photon sources.
\end{abstract}

\maketitle


\section{\label{sec:1} Introduction}
Remote entanglement generation plays a central role in distributed quantum information processing including quantum cryptography~\cite{bennettQuantumCryptographyBells1992,ekertQuantumCryptographyBased1991}, distributed quantum computing~\cite{monroeLargescaleModularQuantumcomputer2014,nickersonFreelyScalableQuantum2014,shimizuSimpleLosstolerantProtocol2025a}, and distributed quantum sensing~\cite{zhangDistributedQuantumSensing2021,liuDistributedQuantumPhase2021,maliaDistributedQuantumSensing2022}. 

\begin{figure}[t]
    \centering
    \includegraphics[width=8.6 truecm]{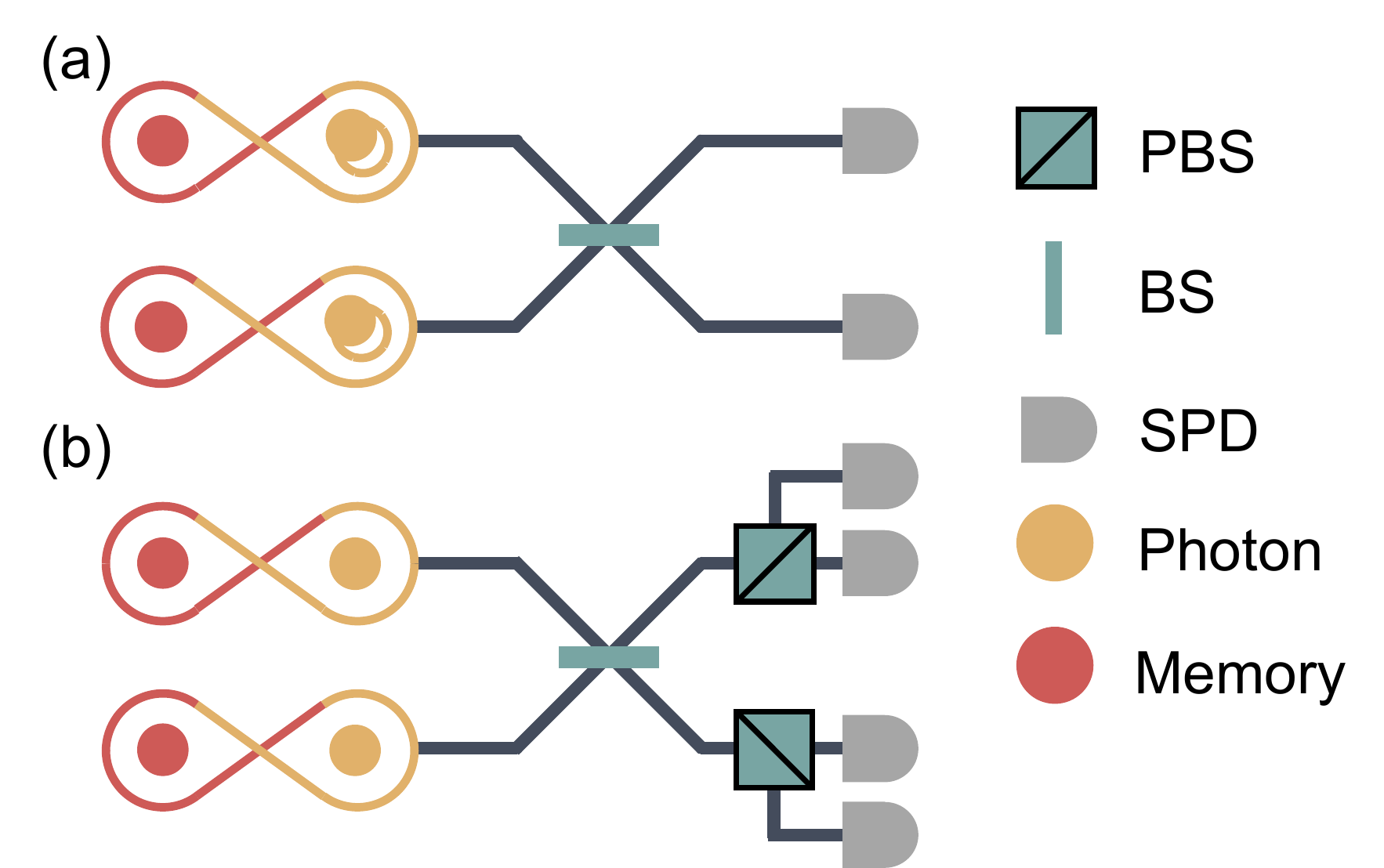}
    \caption{Two types of entanglement swapping protocols. (a) Single-photon protocol (type-I): the photonic qubit is encoded in the vacuum and single-photon states. (b) Two-photon protocol (type-II): the photonic qubit is encoded in polarization states.}
    \label{fig:1}
\end{figure}

Entanglement distribution of qubits has been demonstrated in various physical platforms, including photons~\cite{sunFieldTestEntanglement2017}, color centers in diamond~\cite{knautEntanglementNanophotonicQuantum2024}, solid-state quantum memories~\cite{lago-riveraTelecomheraldedEntanglementMultimode2021}, neutral atoms~\cite{Rosenfeld2017,Hofmann2012,VanLeent2022}, ions~\cite{Hucul2015,Krutyanskiy2023}, and atomic ensembles~\cite{yuEntanglementTwoQuantum2020}. In these experiments, entanglement swapping is employed as a key protocol for distributing entanglement. Figure~\ref{fig:1} shows two typical entanglement swapping protocols. In both protocols, photons entangled with atoms are sent and measured at a central node, where a Bell measurement is performed. As a result, the atom-photon entanglement is swapped into the atom-atom entanglement. Figure~\ref{fig:1}(a) shows a protocol that encodes photonic qubits into vacuum and single-photon states (type-I protocol)~\cite{cabrilloCreationEntangledStates1999}, while Fig.~\ref{fig:1}(b) shows a protocol that encodes photonic qubits into polarizations (type-II protocol)~\cite{zukowskiEventreadydetectorsBellExperiment1993}. Under ideal conditions (lossless channels and deterministic photon sources), the success probability of swapping is 1/2. Considering realistic experimental conditions, including photon generation probability $p$ and channel transmittivity $\eta$, the success probabilities become $\eta p/2$ for the type-I protocol and $\eta^2 p^2 /2$ for the type-II protocol, reflecting differences in the required number of detected photons.

Light has higher number of degrees of freedom (DOFs) beyond two dimensions, such as path~\cite{shimizuQuantumStateEstimation2025,luoQuantumTeleportationHigh2019,huExperimentalHighDimensionalQuantum2020}, time-bin~\cite{ikutaImplementationQuantumState2017,sahaHighfidelityRemoteEntanglement2025}, photon number~\cite{erkilicSurpassingRepeaterlessBound2023}, and orbital angular momentum (OAM)~\cite{groblacherExperimentalQuantumCryptography2006,dingHighdimensionalEntanglementDistant2016}, which can be treated as qudits, high-dimensional quantum systems. Photonic qudits have the potential to enhance the capabilities of quantum communication~\cite{erhardAdvancesHighdimensionalQuantum2020,Yamazaki2025}.
Unfortunately, however, qudit-based Bell measurements cannot be performed solely with linear optics and single-photon detectors~\cite{calsamigliaGeneralizedMeasurementsLinear2002}. Therefore, nonlinear processes or additional resources must be introduced, such as auxiliary single photons~\cite{goyalTeleportingPhotonicQudits2013,goyalQuditTeleportationPhotonsLinear2014,zhangArbitraryTwoparticleHighdimensional2019,zhangQuantumTeleportationPhotonic2019,luoQuantumTeleportationHigh2019,huExperimentalHighDimensionalQuantum2020,paesaniSchemeUniversalHighDimensional2021,baccoProposalPracticalMultidimensional2021,bharosEfficientHighDimensionalEntangled2024a, bhattiHeraldingHigherDimensionalBell2024}, squeezing~\cite{bianchiPredetectionSqueezingResource2025}, or nonlinear crystals~\cite{sephtonQuantumTransportHighdimensional2023a,bianchiNonlinearProtocolHighdimensional2024a}. Protocols employing auxiliary photons generally requires at least $d-2$ photons for $d$-dimensional qudits.

While various high-dimensional Bell measurements have been proposed, protocols using auxiliary photons suffers form a quadratic decrease of the success probability as the dimension increases, which scales as $\mathcal{O}(1/d^2)$. Furthermore, accounting for channel transmittivity and photon generation probability, similar to qubit-based Bell measurements, the success probability decreases exponentially with the scaling of $\mathcal{O}(\eta p)^d/d^2$. In ref.~\cite{baccoProposalPracticalMultidimensional2021}, numerical simulation of three- and four-dimensional entanglement swapping with two-mode squeezed states demonstrated that qudit-based entanglement swapping is advantageous only under specific noise conditions.

In this paper, we propose a high-rate qutrit-based entanglement swapping protocol that surpass conventional two-photon detection-based qubit protocol. Our protocol encodes qutrits into photon-number states combined with an additional mode basis, such as polarization, time-bin, or path. This encoding reduces the number of detected photons, thereby improving the rate of entanglement swapping. We quantify and compare the entanglement generation rate of our protocol to that of type-II protocol using product of entanglement entropy and success probability. We show that our protocol outperforms the type-II protocol, particularly when the photon generation probability is below 0.7. The regime aligns with the typical operational conditions of probabilistic photon sources, such as spontaneous parametric down-conversion and atomic systems (including artificial atoms and ions).

\section{\label{sec:2} Entanglement swapping protocol}

In this section, we describe our qutrit entanglement swapping protocol. 
In Sec.~IIA, we present specific examples of photon sources. In Sec.~IIB, we explain the details of our protocol using qutrits encoded into photon-number and polarization basis. In Sec.~IIC, we analyze the protocol considering practical imperfections, such as photon loss and threshold detection. Finally, in Sec.~IID, we generalize our Bell measurement scheme to arbitrary photonic modes, including path and time-bin. 


\begin{figure}[t]
    \centering
    \includegraphics[width=8.6 truecm]{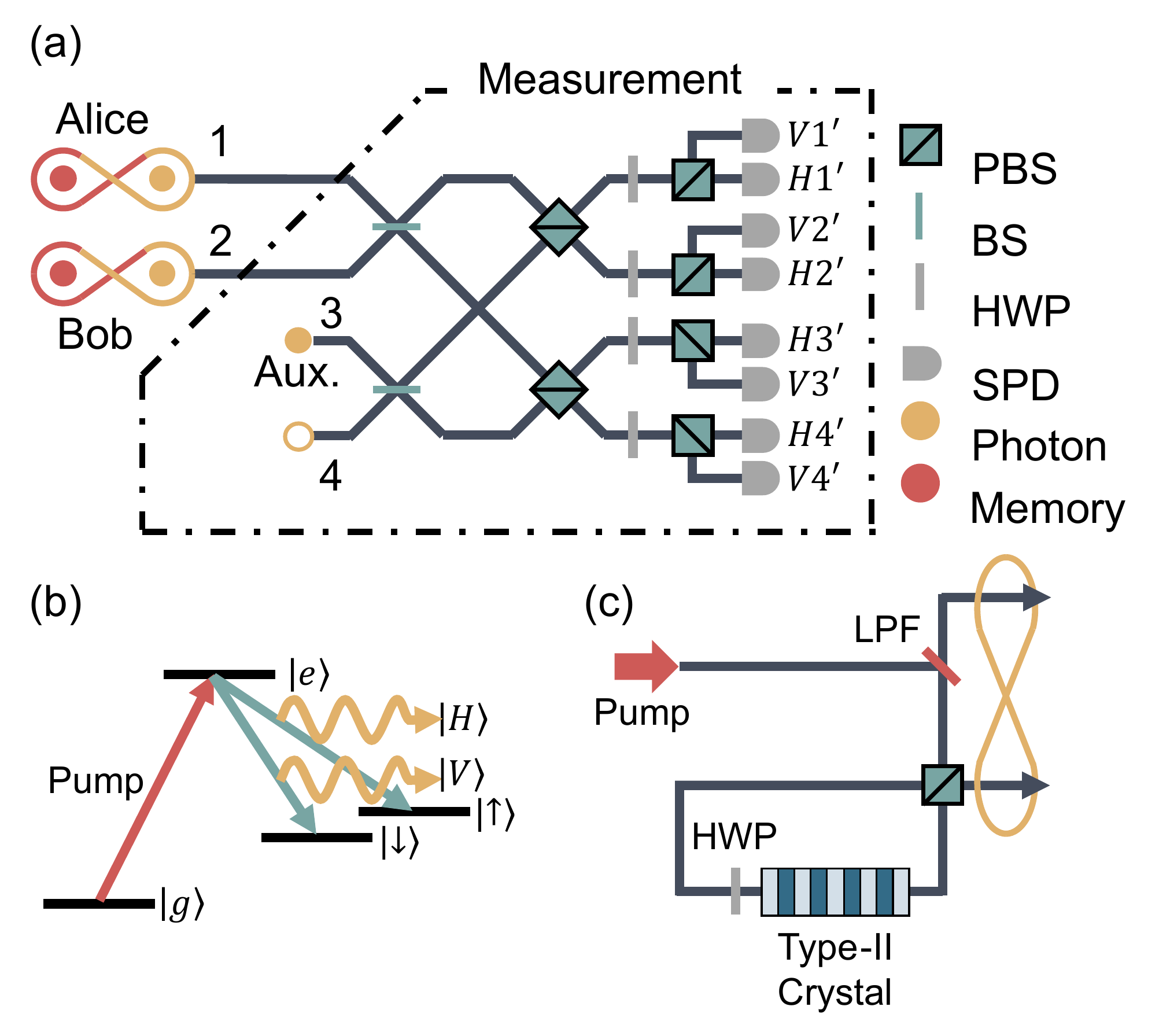}
    \caption{Schematic of the qutrit entanglement swapping protocol. (a) Bell measurement setup. BS: beam splitter; PBS: polarization beam splitter; HWP: half-wave plate; SPD: single-photon detector. (b) Atom–photon qutrit entanglement generated at network nodes using a four-level atom. (c) Photonic qutrit entanglement generated at network nodes using a Sagnac-type SPDC source.}
    \label{fig:2}
\end{figure}

\subsection{\label{sec:2.A}Qutrit Entanglement preparation in network nodes}
Figure~\ref{fig:2}(a) illustrates an example setup of our protocol. In quantum network nodes, Alice and Bob prepare non-maximally entangled qutrit states:
\begin{align}
    \label{eq:1}
    \ket{\Psi}_\mu=\sqrt{1-p}\ket{\bar{0}0}_{\rm mp}+\sqrt{\frac{p}{2}}\left(\ket{\bar{1}H}_{\rm mp}+\ket{\bar{2}V}_{\rm mp}\right),
\end{align}
where the subscripts $\mu=A,\ B$ on the left-hand side denote Alice or Bob, the subscript ${\rm mp}$ on the right-hand side denotes memory-photon entanglement, $p$ is the photon generation probability, $\ket{\Bar{n}}\ (n=0,1,2)$ represents memory qutrit states, and $\ket{0},\ \ket{H},$ and $\ket{V}$ are the vacuum, horizontal polarization, and vertical polarization states of the photonic qutrit, respectively.

Figures~\ref{fig:2}(b) and (c) show two possible implementation of the entanglement described in Eq.~(\ref{eq:1}). The first example, shown in Fig.~\ref{fig:2}(b), is atom-photon entanglement.
Here, the qutrit is encoded in the internal states of four-level atoms, consisting of a ground state $\ket{g}$, an excited state $\ket{e}$, and two different spin states $\ket{\uparrow},\ket{\downarrow}$. Qutrit is encoded in $\ket{g}$, $\ket{\uparrow}$, and $\ket{\downarrow}$. The transition $\ket{e}\leftrightarrow\ket{\uparrow}$ and $\ket{e}\leftrightarrow\ket{\downarrow}$ is coupled to polarized single-photon $\ket{H}$ and $\ket{V}$, respectively. The atom, initially prepared in the ground state $\ket{g}$, is pumped to the excited state $\ket{e}$ by the strong field and subsequently decays to $\ket{\uparrow},\ket{\downarrow}$ along with polarized single-photon emission. This process generates the following qutrit entanglement,
\begin{equation}
    \label{eq:2}
    \ket{\Psi}_{\rm ap}=\sqrt{1-2p}\ket{g0}+\sqrt{p}\left(\ket{\uparrow H}+\ket{\downarrow V}\right),
\end{equation}
where $p$ is the probability of emitting each polarized single photon. Note that, upon post-selecting a photo-emission event, the resulting state becomes a spin-polarization entanglement, $(\ket{\uparrow H}+\ket{\downarrow V})/\sqrt{2}$, which is used for the standard two-photon entanglement swapping~\cite{Krutyanskiy2023}. 
Thus, the atomic system considered here has already been employed in various experiments.

The second example is spontaneous parametric down conversion (SPDC) in Fig.~\ref{fig:2}(c). SPDC with a type-II nonlinear crystal generates a two-mode squeezed vacuum (TMSV) state with different polarization modes:
\begin{equation}
    \label{eq:3}
    \ket{\rm TMSV}=\sqrt{1-\lambda^2}\sum_{n=0}^\infty \lambda^n\ket{nn}_{HV},
\end{equation}
where $\lambda = \tanh r$, and $r$ is a squeezing parameter, and $\ket{n}$ represents a $n$-photon state.
As shown in Fig.~\ref{fig:2}(c), when a crystal is in a Sagnac interferometer, two TMSV states generated from each side of the crystal are interfered at the polarization beamsplitter. Under the condition $\lambda\ll1$, the Fock states with $n\geq2$ can be negligible, and the output state from the Sagnac interferometer is approximated as:
\begin{equation}
    \label{eq:4}
    \ket{\Psi}_{1,2}\approx \sqrt{1-\lambda^2}\left(\ket{0}\ket{0}+\lambda\left(\ket{H}\ket{V}+\ket{V}\ket{H}\right)\right),
\end{equation}
where $\ket{H}$ and $\ket{V}$ are polarization states of single photons, and the subscripts $1$ and $2$ denote the output modes of the Sagnac interferometer. 
Again, $\ket{\Psi}_{1,2}$ is a polarization-entangled state if a photo-emission event is post-selected.
Thus, the setup of the photon source has already been employed in polarization entanglement generation~\cite{Shi2004}.

\subsection{\label{sec:2.B}Photonic Bell measurement for qutrit}
In this section, we describe a photonic Bell measurement for qutrits.
As explained above, each node emits a photonic qutrit encoded into the vacuum state $\ket{0}$ and the polarization states $\ket{H},\ket{V}$. 
To perform a Bell measurement with this encoding, two essential operations must be realized: (i) erasing the "which-node" information of detected single photons, and (ii) projecting onto the superposed states of $\ket{0}$, $\ket{H}$, and $\ket{V}$. 
We design the Bell measurement, as shown in Fig~\ref{fig:2}(a), to fulfill these demands. The BS1 achieves the condition (i), while auxiliary photons and other optics realize (ii). The state of the auxiliary photon is
\begin{equation}
    \label{eq:5}
    \ket{\rm aux}=\sqrt{1-\alpha^2}\ket{0}+\alpha\ket{HV}.
\end{equation}

In the following, we briefly describe our entanglement swapping protocol. For now, we assume ideal conditions, i.e., no photon loss and photon-number-resolving detection (PNRD). Noisy scenarios will be considered in Sec.~\ref{sec:2.C}. The initial state can be expressed as:
\begin{equation}
    \label{eq:6}
    \ket{\Psi_{\rm ini}}=\ket{\Psi}_A\ket{\Psi}_B\ket{\rm aux},
\end{equation}
where $\ket{\Psi}_A$ and $\ket{\Psi}_B$ are defined in Eq.~(\ref{eq:1}).

The initial state is transformed through beamsplitters, polarization beamsplitters and half-wave plates. First, a beamsplitter unitary transforms creation and annihilation operators as follows.
\begin{equation}
    \label{eq:7}
    \begin{bmatrix}
        a_{\mu i'}\\
        a_{\mu j'}
    \end{bmatrix}=\frac{1}{\sqrt{2}}\begin{bmatrix}
        1&1\\
        1&-1
    \end{bmatrix}\begin{bmatrix}
        a_{\mu i}\\
        a_{\mu j}
    \end{bmatrix},
\end{equation}
where the subscript $\mu$ indicates polarization $H$ or $V$, and the subscripts $i,j,i',j'\in\{1,2,3,4\}$ denote the path modes in Fig.~\ref{fig:1}(a). Second, at the polarization beamsplitters, vertically polarized photons pass through, while horizontally polarized photons are reflected. Thus, the unitary matrix transforms the creation and annihilation operators as follows.
\begin{equation}
    \label{eq:8}
    \begin{bmatrix}
        a_{H i'}\\
        a_{H j'}\\
        a_{V i'}\\
        a_{V j'}\\
    \end{bmatrix}=\begin{bmatrix}
        0&1&0&0\\
        1&0&0&0\\
        0&0&1&0\\
        0&0&0&1
    \end{bmatrix}\begin{bmatrix}
        a_{H i}\\
        a_{H j}\\
        a_{V i}\\
        a_{V j}\\
    \end{bmatrix}.
\end{equation}
Finally, the half wave-plates rotate polarizations as follows.
\begin{equation}
    \label{eq:9}
    \begin{bmatrix}
        a_{H i'}\\
        a_{V i'}
    \end{bmatrix}=\frac{1}{\sqrt{2}}\begin{bmatrix}
        1&1\\
        1&-1
    \end{bmatrix}\begin{bmatrix}
        a_{H i}\\
        a_{V i}
    \end{bmatrix}.
\end{equation}
By combining these transformations, we obtain the overall input-output transformation matrix for annihilation operators $a_{\mu i}$ (input) and $\ b_{\mu i'}$ (output) with $\mu=H,V$ and $i=1,2,3,4$.
\begin{widetext}
    \begin{equation}
    \label{eq:10}
        \begin{bmatrix}
            b_{H1'}\\
            b_{H2'}\\
            b_{H3'}\\
            b_{H4'}\\
            b_{V1'}\\
            b_{V2'}\\
            b_{V3'}\\
            b_{V4'}\\
        \end{bmatrix}=\frac{1}{2}
        \begin{bmatrix}
            0&0&1&1&1&1&0&0\\
            1&1&0&0&0&0&1&1\\
            0&0&1&-1&1&-1&0&0\\
            1&-1&0&0&0&0&1&-1\\
            0&0&1&1&-1&-1&0&0\\
            1&1&0&0&0&0&-1&-1\\
            0&0&1&-1&-1&1&0&0\\
            1&-1&0&0&0&0&-1&1\\
        \end{bmatrix}\begin{bmatrix}
            a_{H1}\\
            a_{H2}\\
            a_{H3}\\
            a_{H4}\\
            a_{V1}\\
            a_{V2}\\
            a_{V3}\\
            a_{V4}\\
        \end{bmatrix}.
    \end{equation}
\end{widetext}
After these transformations, photons are detected by the single photon detectors. For example, we consider a scenario in which single photons are detected at $H1'$ and $H2'$. The resulting heralded atomic entangled state is
\begin{eqnarray}
    \label{eq:11}
    \ket{\Phi}_{H1'H2'}&&=\frac{(1-p)\alpha}{4\sqrt{P_{\rm s}}}\ket{\Bar{0}\Bar{0}}\nonumber\\
    &&+\frac{p\sqrt{1-\alpha^2}}{8\sqrt{P_{\rm s}}}\left(\ket{\Bar{1}\Bar{2}}+\ket{\Bar{2}\Bar{1}}\right),
\end{eqnarray}
where $P_{\rm s}$ is the success probability for generating entanglement. To maximize the fidelity between this state and maximally entangled qutrit state, which is defined as,
\begin{eqnarray}
    \label{eq:12}
    \ket{\rm Bell}=\left(\ket{\Bar{0}\Bar{0}}+\ket{\Bar{1}\Bar{2}}+\ket{\Bar{2}\Bar{1}}\right)/\sqrt{3},
\end{eqnarray}
the coefficients in Eq.~(\ref{eq:11}) should be equal, yielding the relationship between parameters $\alpha$ and $p$:
\begin{eqnarray}
    \label{eq:13}
    \alpha=\frac{p}{\sqrt{5p^2-8p+4}}.
\end{eqnarray}
Substituting this relationship into Eq.~(\ref{eq:11}) results in the maximally entangled qutrit state $\ket{\rm Bell}$ with the success probability:
\begin{equation}
    \label{eq:14}
    P_{\rm s, H1'H2'}=\frac{3}{16}\frac{p^2(1-p)^2}{5p^2-8p+4}
\end{equation}

There are additional detection patterns heralding entanglement generation.
These patterns and corresponding entangled states are summarized in Table~\ref{tab:1}. Consequently, the total success probability is
\begin{equation}
    \label{eq:15}
    P_{\rm s}=\frac{3p^2(1-p)^2}{5p^2-8p+4}.
\end{equation}
\begin{table}[tbp]
    \centering
    \caption{Detection patterns of single photons and heralded entangled states.}
    \begin{ruledtabular}
    \begin{tabular}{@{\hspace{2em}} cc @{\hspace{2em}}}
    Detection patterns & Heralded entanglement \\ \hline 
    \raisebox{2.8ex}{}$(H1',H2')$& $\left(\ket{\bar{0}\bar{0}}+\ket{\bar{1}\bar{2}}+\ket{\bar{2}\bar{1}}\right)/\sqrt{3}$\\
    $(H1',V2')$&$\left(-\ket{\bar{0}\bar{0}}+\ket{\bar{1}\bar{2}}+\ket{\bar{2}\bar{1}}\right)/\sqrt{3}$\\
    $(V1',H2')$&$\left(-\ket{\bar{0}\bar{0}}+\ket{\bar{1}\bar{2}}+\ket{\bar{2}\bar{1}}\right)/\sqrt{3}$\\
    $(V1',V2')$& $\left(\ket{\bar{0}\bar{0}}+\ket{\bar{1}\bar{2}}+\ket{\bar{2}\bar{1}}\right)/\sqrt{3}$\vspace{0.5ex}\\\hline
    \raisebox{2.8ex}{}$(H1',H4')$& $\left(\ket{\bar{0}\bar{0}}+\ket{\bar{1}\bar{2}}-\ket{\bar{2}\bar{1}}\right)/\sqrt{3}$\\
    $(H1',V4')$& $\left(\ket{\bar{0}\bar{0}}-\ket{\bar{1}\bar{2}}+\ket{\bar{2}\bar{1}}\right)/\sqrt{3}$\\
    $(V1',H4')$& $\left(\ket{\bar{0}\bar{0}}-\ket{\bar{1}\bar{2}}+\ket{\bar{2}\bar{1}}\right)/\sqrt{3}$\\
    $(V1',V4')$& $\left(\ket{\bar{0}\bar{0}}+\ket{\bar{1}\bar{2}}-\ket{\bar{2}\bar{1}}\right)/\sqrt{3}$\vspace{0.5ex}\\\hline
    \raisebox{2.8ex}{}
    $(H3',H2')$& $\left(\ket{\bar{0}\bar{0}}-\ket{\bar{1}\bar{2}}+\ket{\bar{2}\bar{1}}\right)/\sqrt{3}$\\
    $(H3',V2')$& $\left(\ket{\bar{0}\bar{0}}+\ket{\bar{1}\bar{2}}-\ket{\bar{2}\bar{1}}\right)/\sqrt{3}$\\
    $(V3',H2')$& $\left(\ket{\bar{0}\bar{0}}+\ket{\bar{1}\bar{2}}-\ket{\bar{2}\bar{1}}\right)/\sqrt{3}$\\
    $(V3',V2')$& $\left(\ket{\bar{0}\bar{0}}-\ket{\bar{1}\bar{2}}+\ket{\bar{2}\bar{1}}\right)/\sqrt{3}$\vspace{0.5ex}\\\hline
    \raisebox{2.8ex}{}$(H3',H4')$& $\left(-\ket{\bar{0}\bar{0}}+\ket{\bar{1}\bar{2}}+\ket{\bar{2}\bar{1}}\right)/\sqrt{3}$\\
    $(H3',V4')$& $\left(\ket{\bar{0}\bar{0}}+\ket{\bar{1}\bar{2}}+\ket{\bar{2}\bar{1}}\right)/\sqrt{3}$\\
    $(V3',H4')$& $\left(\ket{\bar{0}\bar{0}}+\ket{\bar{1}\bar{2}}+\ket{\bar{2}\bar{1}}\right)/\sqrt{3}$\\
    $(V3',V4')$& $\left(-\ket{\bar{0}\bar{0}}+\ket{\bar{1}\bar{2}}+\ket{\bar{2}\bar{1}}\right)/\sqrt{3}$\\
    \end{tabular}
    \end{ruledtabular}
    \label{tab:1}
\end{table}

In summary, we realize a photonic Bell measurement for qutrits encoded in polarization and photon-number basis with linear optics, the auxiliary photon, and photon-number-resolving detectors.

\subsection{\label{sec:2.C} Effects of loss and threshold detection}
In this section, we analyze the effects of photon loss and threshold detection which only distinguishes zero and more than one photons.
When all photons to be measured pass through channels with transmittivity $\eta$, the initial state in Eq.~(\ref{eq:6}) is replaced by the following density matrix,
\begin{equation}
    \label{eq:16}
    \rho_{\rm in}^{\rm loss} = \rho_A^{\rm loss}\otimes\rho_B^{\rm loss} \otimes \rho_{\rm aux}^{\rm loss},
\end{equation}
where the states from Alice and Bob are given by
\begin{align}
    \label{eq:17}
    \rho_\mu&=\ket{\Psi^{(0)}}\bra{\Psi^{(0)}}_\mu+\rho_\mu^{\rm (1)},\\
    \label{eq:18}
    \ket{\Psi^{(0)}}_\mu&=\sqrt{1-p}\ket{\Bar{0}0}
    +\sqrt{\eta p}(\ket{\Bar{1}H}+\ket{\Bar{2}V}),\\
    \label{eq:19}
    \rho_\mu^{\rm (1)}&=\frac{p}{2}(1-\eta)\left(\ket{\Bar{1}}\bra{\Bar{1}}+\ket{\Bar{2}}\bra{\Bar{2}}\right)\otimes\ket{\rm 0}\bra{\rm 0},
\end{align}
with $\mu=A,B$. The state of the auxiliary photons becomes
\begin{align}
    \label{eq:20}
    \rho_{\rm aux}^{\rm loss}&=\ket{\rm aux^{(0)}}\bra{\rm aux^{(0)}}+\rho_{\rm aux}^{(1)},\\
    \label{eq:21}
    \ket{\rm aux^{(0)}}&=\sqrt{1-\abs{\alpha}^2}+\alpha \eta \ket{HV},\\
    \label{eq:22}
    \rho_{\rm aux}^{(1)}&=\abs{\alpha}^2\sqrt{\eta(1-\eta)}\left(\ket{H0}\bra{H0}+\ket{0V}\bra{0V}\right)\nonumber\\
    &+\abs{\alpha}^2(1-\eta)^2\ket{00}\bra{00}
\end{align}

The threshold detectors perform the following measurement operators.
\begin{eqnarray}
    \label{eq:23}
    &&\hat{M_0}=\ket{0}\bra{0}\\
    \label{eq:24}
    &&\hat{M_1}=\mathbf{1}-\ket{0}\bra{0},
\end{eqnarray}
where $\mathbf{1}$ is the identity operator.

We consider the same entanglement swapping calculation in Sec.~\ref{sec:2.B}, but replace the initial state with $\rho_{\rm ini}^{\rm loss}$ and the measurements with $\hat{M}_0$ and $\hat{M}_1$. Using these measurement operators, we define measurement operators corresponding to the measurement block enclosed by the dash-dotted lines in Fig~\ref{fig:2}(a). Let $M_{\rm th}$ denote the measurement operator for threshold detector, while $M_{\rm pnr}$ is the measurement operator for PNRD.
For example, when the photons are detected in modes $H1'$ and $H2'$, the unnormalized post-measurement state under threshold detection is 
\begin{align}
    \rho_{\rm out}^{\rm th}=\Tr_{M}\left[\hat{M}_{\rm th}^{H1'H2'} \rho_{\rm in}^{\rm loss}\right].
\end{align}
Under PNRD, we replace $M_1$ with $\ketbra{1}{1}$ and obtain 
\begin{align}
    \rho_{\rm out}^{\rm PNR}=\Tr_{M}\left[\hat{M}_{\rm PNR}^{H1'H2'} \rho_{\rm in}^{\rm loss}\right],
\end{align}
where the measurement operator $M_{\rm th}^{H1'H2'}$ and $M_{\rm PNR}^{H1'H2'}$ are derived in appendix~\ref{adx:A}, and $\Tr_M$ denotes partial trace over the measurement modes in Fig.~\ref{fig:2}(a). In the derivation of these measurement operators, we used the description of path-mode representation for simplicity, as shown in Fig~\ref{fig:3}.

The corresponding success probabilities are
\begin{align}
    P_{s}^{\rm th}&=16\Tr \rho_{\rm out}^{\rm th}\\
    P_{s}^{\rm PNR}&=16\Tr \rho_{\rm out}^{\rm PNR}.
\end{align}
and the fidelities with respect to the maximally entangled qutrit state $\ket{\rm Bell}$ are given by
\begin{align}
    F_{\rm th}&=\left(\Tr \rho_{\rm out}^{\rm th}\right)^{-1}\bra{\rm Bell} \rho_{\rm out}^{\rm th} \ket{\rm Bell}\\
    F_{\rm PNR}&=\left(\Tr \rho_{\rm out}^{\rm PNR}\right)^{-1}\bra{\rm Bell} \rho_{\rm out}^{\rm PNR} \ket{\rm Bell}.
\end{align}

\subsection{\label{sec:2.D} Generalization to arbitral modes}
\begin{figure}
    \centering
    \includegraphics[width=8.6 truecm]{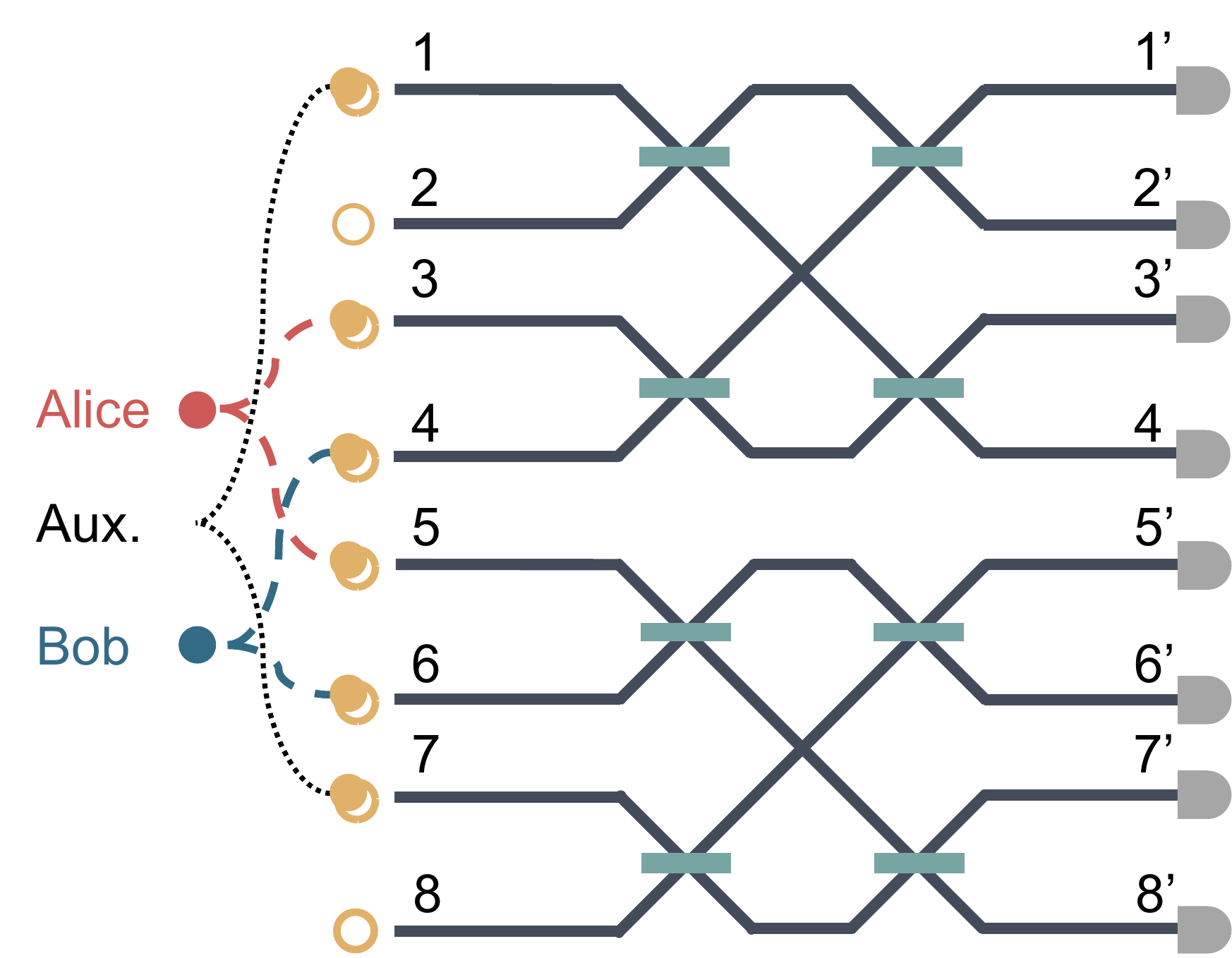}
    \caption{Implementation of the qutrit Bell measurement using path-mode encoding. The circuit performs a four-dimensional Hadamard transformation on each mode group, enabling projection onto the qutrit Bell basis.}
    \label{fig:3}
\end{figure}

In the previous subsections, we described our protocol with polarization as the mode basis. Here, we generalize our Bell measurement scheme to arbitrary photonic modes, such as path or time-bin degrees of freedom. By reordering the input-output transformation matrix in Eq.~(\ref{eq:10}), it is given by the following block-diagonalized matrix $U$:
\begin{align}
    \label{eq:25}
    U&=\begin{bmatrix}
        H_4 & 0\\
        0 & H_4
    \end{bmatrix}\\
    \label{eq:26}
    H_4&=\begin{bmatrix}
        1&1&1&1\\
        1&-1&1&-1\\
        1&1&-1&-1\\
        1&-1&-1&1
    \end{bmatrix},
\end{align}
where $H_4$ is a four-dimensional Hadamard matrix.
The corresponding reordering of annihilation operators is
\begin{align}
    \label{eq:27}
    &\left[ a_{\rm H1},a_{\rm H2},a_{\rm H3},a_{\rm H4},a_{\rm H5},a_{\rm H6},a_{\rm H7},a_{\rm H8}\right]^T\nonumber\\
    &\rightarrow\left[a_{\rm H3},a_{\rm H4},a_{\rm V1},a_{\rm V2},a_{\rm H1},a_{\rm H2},a_{\rm V3},a_{\rm V4}\right]^T\\
    \label{eq:28}
    &\left[ b_{\rm H1'},b_{\rm H2'},b_{\rm H3'},b_{\rm H4'},b_{\rm H5'},b_{\rm H6'},b_{\rm H7'},b_{\rm H8'}\right]^T\nonumber\\
    &\rightarrow\left[b_{\rm H1'},b_{\rm H3'},b_{\rm V1'},b_{\rm V3'},b_{\rm H2'},b_{\rm H4'},b_{\rm V2'},b_{\rm V4'}\right]^T
\end{align}
For notational simplicity, we relabel the subscripts in Eqs.~(\ref{eq:27}) and (\ref{eq:28}) as 1 through 8 in the following derivation, i.e., $\left[a_1,a_2,\ldots,a_8\right]^T$ and $\left[b_{1'},b_{2'},\ldots,b_{8'}\right]^T$.

Figure~\ref{fig:3} depicts the interferometer implementing Eq.~(\ref{eq:25}), composed of two independent 4-input, 4-output interferometers that realize the Hadamard transformation. In this general framework, the entangled states prepared by Alice and Bob are
\begin{align}
    \label{eq:29}
    \ket{\Psi'}_A&=\sqrt{1-p}\ket{\Bar{0}}\ket{0}+\sqrt{\frac{p}{2}}\left(\ket{\Bar{1}}\ket{1}_3+\ket{\Bar{2}}\ket{1}_5\right)\\
    \label{eq:30}
    \ket{\Psi'}_B&=\sqrt{1-p}\ket{\Bar{0}}\ket{0}+\sqrt{\frac{p}{2}}\left(\ket{\Bar{1}}\ket{1}_4+\ket{\Bar{2}}\ket{1}_6\right)\\
    \label{eq:31}
    \ket{\rm aux'}&=\sqrt{1-\abs{\alpha}^2}\ket{0}+\alpha a_1^\dagger a_7^\dagger\ket{1}_1\ket{1}_7,
\end{align}
where $\ket{\Bar{n}}\ (n=0,1,2)$ represents the memory qutrit states as in Eq~(\ref{eq:1}), $\ket{0}$ is the vacuum state over all modes, and $\ket{1}_k$ is a single-photon state in mode $k$ with all other modes in the vacuum state. 

In this description, the photonic qutrit in Alice's (Bob's) node is encoded into vacuum, $\ket{1}_{3}$ ($\ket{1}_{5}$), and $\ket{1}_{4}$ ($\ket{1}_{6}$), i.e., in the path degree of freedom. To apply this to time-bin encoding, we replace $\ket{1}_{3}$ ($\ket{1}_{5}$) and $\ket{1}_{4}$ ($\ket{1}_{6}$) with $\ket{\rm early}_{A(B)}$ and $\ket{\rm late}_{A(B)}$ where $\ket{\rm early}$ and $\ket{\rm late}$ are the first and second time-bin. In this encoding, the interferometer is the 4-input, 4-output Hadamard transformation
Therefore, the proposed protocol is directly applicable to a variety of photonic mode encodings, including time-bin and path, by adapting the structure of interferometers to mach the mode basis.

\section{\label{sec:3}Result}

In this section, we evaluate the performance of our qutrit entanglement swapping protocol. We compare it to the type-II qubit-based entanglement swapping protocol under both ideal and non-ideal conditions. These two protocols are directly comparable because both rely on the detection of two photons. We also evaluate the success probability and fidelity of qutrit entanglement swapping protocol in the presence of imperfections.

We first consider the entanglement generation rate in the ideal case. As a measure of entanglement, we use the von Neumann entanglement entropy~\cite{nielsenQuantumComputationQuantum2012} since the output state is pure. For a pure state $\ket{\Psi}_{AB}$ on bipartite system $AB$, its entanglement entropy is defined as below.
\begin{align}
    S(\rho_A)=-\Tr[\rho_A \log_2 (\rho_A)],
\end{align}
where $\rho_A=\Tr_B[\ket{\psi}\bra{\psi}_{AB}]$ is a reduced density matrix given by taking partial trace for system $B$. Note that $S(\rho_A)=S(\rho_B)$, and the entanglement entropy reaches its maximum value $\log_2 d$ for maximally entangled states in $d$-dimensional systems. We define the entanglement generation rate as the product $S(\rho_A)\times P_s$, where $P_s$ is the success probability of the protocol

\begin{figure}
    \centering
    \includegraphics[width=8.6 truecm]{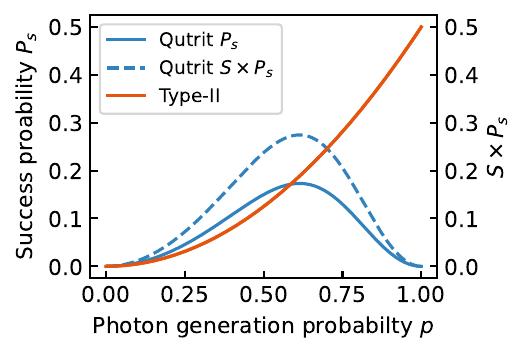}
    \caption{Success probability and entanglement generation rates for photon generation probability without imperfections. Blue solid line: Success probability of qutrit-based entanglement swapping (left-vertical axis). Blue dashed line: entanglement generation rate (right-vertical axis). Orange solid line: both success probability and entanglement generation rate of type-II qubit-based entanglement swapping.}
    \label{fig:4}
\end{figure}

Figure~\ref{fig:4} shows the success probability $P_s$ and the entanglement generation rate $S(\rho_A)\times P_s$ as functions of the photon generation probability $p$. The blue solid line represents success probability of the qutrit protocol, while the blue dashed line shows the corresponding entanglement generation rate, given by $\log_2 (3)\times P_s$. The orange solid line represents both the success probability and the entanglement generation rate of the type-II qubit protocol, which are equal to $p^2/2$ since its entanglement entropy is 1. 

The maximum success probability of our protocol is
\begin{align}
    \label{eq:33}
    P_{\rm tot}&=\frac{3}{125}\frac{16+97\sqrt[3]{2}-76\sqrt[3]{4}}{-2+\sqrt[3]{2}+2\sqrt[3]{2}}\approx0.173
\end{align}
when
\begin{align}
    p&=\frac{4-2\sqrt[3]{2}+\sqrt[3]{4}}{5}\approx0.614.
\end{align}
The entanglement generation rate is also maximized at this photon generation probability. The crossover point where the qutrit protocol surpasses the qubit protocol occurs at approximately $p\approx 0.701$. Therefore, our protocol provides a higher entanglement generation rate than the type-II protocol when the photon generation probability is below 0.7.

\begin{figure}
    \centering
    \includegraphics[width=8.6 truecm]{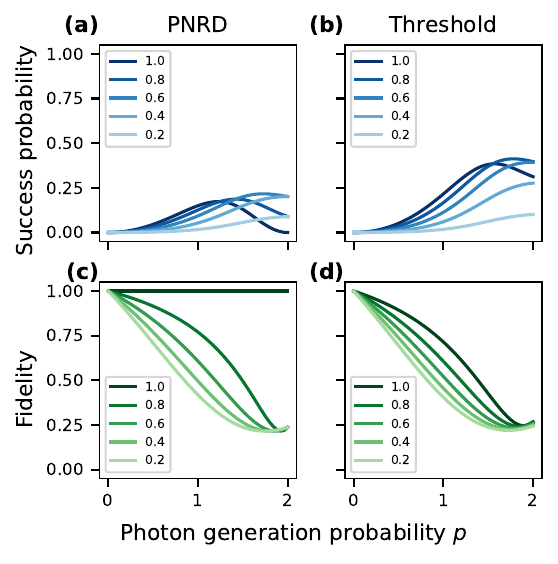}
    \caption{Success probability and fidelity of the qutrit-based protocol as functions of the photon generation probability $p$, under imperfect conditions. (a) Success probability with photon-number-resolving detectors (PNRD). (b) Success probability with threshold detectors. (c) Fidelity of the generated qutrit entangled state with PNRD. (d) Fidelity with threshold detectors. Darker colors represent higher channel transmittivity $\eta$, while lighter colors indicate lower $\eta$.}
    \label{fig:6}
\end{figure}

Next, we evaluate the success probability and the fidelity under imperfections as shown in Fig.~\ref{fig:6}. Figures~\ref{fig:6}(a) and (b) show the success probability versus photon generation probability for PNRD and threshold detection. Figures.~\ref{fig:6}(c) and (d) show the corresponding fidelity with respect to the maximally entangled qutrit state. These values are calculated by Eq.~(27)-(30). The darker colors correspond higher transmittivity. 

In the low-transmittivity regime, the success probability may be higher than the ideal case due to the contribution of undesired input states, while the fidelity decreases with increasing loss. We find that there are no huge differences of fidelities between PNRD and threshold detection $\eta<0.8$. This is because, in our protocol, the number of incident photons is at most four, and due to the probabilistic nature of photon generation as well as channel loss, the number of photons that actually reach the detectors is typically small. To maintain high fidelity in the presence of loss, lower photon generation probabilities are preferable. This condition is compatible with the characteristics of the photon sources introduced in Fig.~\ref{fig:2}(b) and (c). Atomic sources suffer from spontaneous emission noise, which can be mitigated with pump pulses shorter than the life-time of excited states~\cite{tanjiRatefidelityTradeoffCavitybased2024a}, naturally leading a small photon generation probability. Similarly, a SPDC photon source must be operated in sufficiently low photon generation probability to suppress multi-photon effects.

\section{Conclusion}
In this paper, we proposed a high-rate entanglement swapping protocol for qutrits. By encoding qutrits in a hybrid basis consisting of photon-number states and an additional mode, such as polarization, we demonstrated that our protocol achieves a higher entanglement generation rate than conventional two-photon detection-based qubit protocols.

We considered photonic qutrits encoded in vacuum and polarization states, which are suitable for various photon sources, including single atoms and SPDC-based photon sources. Furthermore, our scheme can be generalized to other photonic degrees of freedom, such as path and time-bin.

We evaluated entanglement generation rate as the product of the success probability and an entanglement measure, under both ideal and non-ideal conditions, including photon loss and threshold detection. We showed that our protocol outperforms the type-II qubit-based protocol across all transmittivity levels when the photon generation probability is below approximately 0.7.

We also analyzed the success probability and fidelity under imperfections. We found higher fidelity can still be achieved under significant photon loss by reducing photon generation probability. This feature aligns with the characteristics of realistic photon sources: atomic sources tend to operate at low photon generation rates due to spontaneous emission noise, and SPDC sources must be driven in the low-gain regime to suppress multi-photon contamination.

In this study, we focused on comparison with the type-II protocol, which, like ours, relies on two-photon detection. However, type-I protocols are known to achieve even higher entanglement generation rates. Designing a qutrit-based entanglement swapping scheme that surpasses type-I performance remains open. Nevertheless, our protocol represents a significant milestone: it is the first qutrit-based scheme to outperform a well-established qubit-based protocol in terms of entanglement generation rate. Extending this approach to arbitrary $d$-dimensional systems while maintaining practical feasibility is a compelling direction for future work.

\begin{acknowledgments}
The authors would like to thank L. Bianchi and K. Azuma for helpful comments on the paper. This work is supported by JST Moonshot R\&D, Grant
Number JPMJMS2061, JST ASPIRE, Grant Number JPMJAP2427, JST SPRING, Grant Number JPMJSP2123, and Program for the Advancement of Next Generation Research Projects, Keio University.
\end{acknowledgments}

\appendix

\section{\label{adx:A}Derivation of measurement operators}
\subsection{Measurement operator for PNRD}
In this section, we derive the measurement operator for the case when photons are detected in detectors $1'$ and $5'$ in Fig.~\ref{fig:3}. The corresponding projection operator for this coincidence detection is
\begin{align}
    \label{eq:A1}
    P_{\rm 1'5'}=\ket{11}\bra{11}_{1'5'}=b_{1'}^\dagger b_{5'}^\dagger\ket{0}\bra{0}b_{1'} b_{5'}.
\end{align}
The projector is transformed by the interferometer unitary $U_{\rm BS}$ in Eq~(\ref{eq:25}).
\begin{align}
    \label{eq:A2}
    \hat{M}_{\rm PNR}&=U_{\rm BS}P_{\rm 1'5'}U_{\rm BS}^\dagger\nonumber\\
    &= \frac{1}{16}\sum_{i,k=1}^{4}\sum_{j,l=5}^{8}a_i^\dagger a_j^\dagger\ket{0}\bra{0}a_ka_l
\end{align}
For the qutrits encoded into photon-number and polarization, the annihilation operators $\left[a_1,a_2,...,a_8\right]$ correspond to:
\begin{align}
    \label{eq:A3}
    &\left[a_{\rm H3},a_{\rm H4},a_{\rm V1},a_{\rm V2},a_{\rm H1},a_{\rm H2},a_{\rm V3},a_{\rm V4}\right]\nonumber\\
    &\leftrightarrow \left[a_1,a_2,...,a_8\right]
\end{align}
This identification allows us to express the measurement operator $\hat{M}_{\rm PNR}$ in terms of photon-number and polarization modes.
\subsection{Measurement operator for threshold detection}
In the case of threshold detection, the detectors cannot resolve the exact number o photons. Therefore, when all photons are detected in mode $1'$ and $5'$, the relevant projection operators must include all possible photon-number combinations. Assuming a maximum of four photons in the system, we introduce the following projectors:
\begin{align}
    \label{eq:A7}
    P_{0}&=\ket{11}\bra{11}_{1'5'}=b_{1'}^\dagger b_{5'}^\dagger\ket{0}\bra{0}b_{1'} b_{5'}\\
    \label{eq:A8}
    P_{1}&=\ket{21}\bra{21}_{1'5'}=\frac{1}{2}b_{1'}^{\dagger2} b_{5'}^{\dagger}\ket{0}\bra{0}b_{1'}^{2} b_{5'}^{}\\
    \label{eq:A9}
    P_{2}&=\ket{12}\bra{12}_{1'5'}=\frac{1}{2}b_{1'}^{\dagger} b_{5'}^{\dagger2}\ket{0}\bra{0}b_{1'}^{} b_{5'}^{2}\\
    \label{eq:A10}
    P_{3}&=\ket{31}\bra{31}_{1'5'}=\frac{1}{6}b_{1'}^{\dagger3} b_{5'}^{\dagger}\ket{0}\bra{0}b_{1'}^{3} b_{5'}^{}\\
    \label{eq:A11}
    P_{4}&=\ket{13}\bra{13}_{1'5'}=\frac{1}{6}b_{1'}^{\dagger} b_{5'}^{\dagger3}\ket{0}\bra{0}b_{1'}^{} b_{5'}^{3}\\
    \label{eq:A12}
    P_{5}&=\ket{22}\bra{22}_{1'5'}=\frac{1}{4}b_{1'}^{\dagger2} b_{5'}^{\dagger2}\ket{0}\bra{0}b_{1'}^{2} b_{5'}^{2}
\end{align}

Thus, the full measurement operator for threshold detection is
\begin{align}
    \hat{M}_{\rm th}=U_{\rm BS}\sum_{i=0}^5 P_i U_{\rm BS}.
\end{align}
Using the mode correspondence in Eq.~(\ref{eq:A3}), this measurement operator can be expressed in terms of photon-number and polarization modes.

\bibliography{ref}

\end{document}